\documentclass[12pt]{iopart}

\usepackage{iopams}
\usepackage[numbers]{natbib}
\usepackage{graphicx}

\begin{document}

\title[Quantum dot occupation and electron dwell time in the cotunneling regime]{Quantum dot occupation and electron dwell time in the cotunneling regime}

\author{B. K\"ung$^{1,2}$, C. R\"ossler$^{1}$, M. Beck$^{3}$, J. Faist$^{3}$, T. Ihn$^{1}$, and K. Ensslin$^{1}$}

\address{$^{1}$Solid State Physics Laboratory, ETH Zurich,
8093 Zurich, Switzerland \newline
$^{2}$Institut N\'eel, CNRS and Universit\'e Joseph Fourier, BP 166, 38042 Grenoble Cedex 9, France\newline
$^{3}$Institute for Quantum Electronics, ETH Zurich, 8093 Zurich, Switzerland}
\ead{bruno.kung@grenoble.cnrs.fr}

\begin{abstract}
We present comparative measurements of the charge occupation and conductance of a GaAs/AlGaAs quantum dot. The dot charge is measured with a capacitively coupled quantum point contact sensor. In the single-level Coulomb blockade regime near equilibrium, charge and conductance signals are found to be proportional to each other. We conclude that in this regime, the two signals give equivalent information about the quantum dot system. Out of equilibrium, we study the inelastic-cotunneling regime. We compare the measured differential dot charge with an estimate assuming a dwell time of transmitted carriers on the dot given by $h/E$, where $E$ is the blockade energy of first-order tunneling. The measured signal is of a similar magnitude as the estimate, compatible with a picture of cotunneling as transmission through a virtual intermediate state with a short lifetime.
\end{abstract}

\maketitle

\section{Introduction}
Quantum dots (QDs) coupled to source and drain electrodes represent versatile and well-controlled systems for the study of mesoscopic transport \cite{Kouwenhoven97}. The many aspects of electron tunneling through QDs are typically studied by measuring either the QD \emph{conductance}, or the QD \emph{charge occupation}. There are several techniques available for measuring the QD charge occupation, among them direct capacitance measurements \cite{Ashoori92a,Ashoori92b,Gabelli06}, and the use of single-electron transistor \cite{Lafarge91,Berman97,Duncan99} and quantum point-contact \cite{Field93} (QPC) electrometers. More insight can be gained when \emph{combining} charge and conductance measurements, and extracting information from both of them to obtain a more complete picture of the system. In the sequential tunneling regime of the QD, combined charge and conductance measurements can be used to determine the system timescales \cite{Schleser05}, whereas in a strong-coupling regime, such measurements reveal the effect of Kondo correlations on the charge of a QD \cite{Sprinzak02}.

In this paper, we present combined transport and QPC charge detection measurements in the cotunneling regime of a QD \cite{Averin90,Glattli91,Franceschi01,Wegewijs01,Zumbuhl04}. Cotunneling is a second-order transmission process through the QD dominating when first-order tunneling is energetically forbidden due to Coulomb blockade. Both elastic \cite{Glattli91,Foxman93} and inelastic \cite{Franceschi01,Zumbuhl04} versions of this process have been studied experimentally. In particular, their coherence properties have been of interest \cite{Sigrist06,Gustavsson08c}. In that context, the general presumption is that environmental decoherence should be weak due to the short duration of cotunneling. The short duration is in turn explained by the large degree of energy uncertainty of the order of $E$, the blockade energy of first-order tunneling. A direct measurement of the cotunneling time is, however, not available. The present paper is based on the idea that the cotunneling time must affect the occupation of the QD, and thus the QPC charge detection measurement should provide information about it. 

Based on this indirect approach, we test the assumption that the cotunneling time is limited by $h/E$. Our method is a quantitative comparison of the measured differential QD charge in the inelastic cotunneling regime with an estimate based on the measured QD current and a carrier dwell time of $h/E$. We find that the measured signal is of the same order of magnitude as the estimate, consistent with a cotunneling time bounded by the time $h/E$. We contrast these results with measurements of the QD charge in a regime where inelastic cotunneling is accompanied by sequential tunneling (cotunneling-assisted sequential tunneling \cite{Wegewijs01,Golovach04,Schleser05cast,Aghassi08}, CAST). The charge signal in this regime is significantly larger than what would be expected assuming a tunneling dwell time of $h/E$. We attribute this to the comparatively long dwell time of sequential tunneling events.

In addition to inelastic cotunneling occurring at nonzero source--drain voltage, we study resonant tunneling at zero source--drain voltage \cite{Averin90,Glattli91,Foxman93}. The finite nonactivated QD conductance in the tails of a resonant peak (in the Coulomb-blockade valley) is termed elastic cotunneling. In this regime, the direct current through the QD vanishes and we demonstrate that the main contribution to the QD charge is due to its equilibrium occupation, unlike in the inelastic regime where the time-averaged charge is mainly due to the dwell time of transmitted carriers. The differential QD conductance and the differential QPC signal are then both a probe of the spectral density of the QD state and are found to agree over two orders of magnitude.

\begin{figure}
\includegraphics{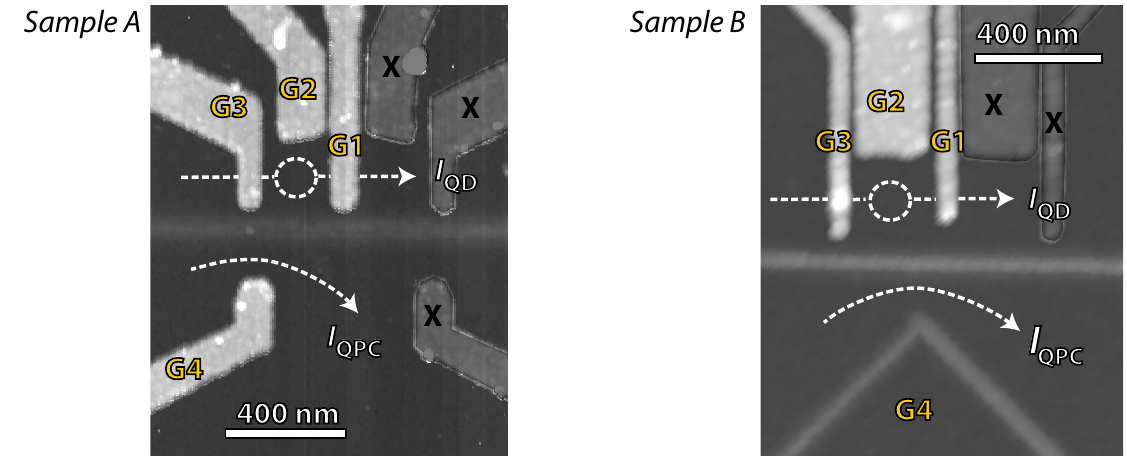}
\caption{Atomic-force micrographs of the two samples used in the experiments. (Sample A: Figures \ref{fig:PS_figure_lowCoupling}, \ref{fig:PS_figure_hiCoupling}, \ref{fig:PS_figure_inelCotunneling}, \ref{fig:PS_figure_DirectCurrentAndTC}. Sample B: Figure \ref{fig:PS_figure_InCo2}.) The dark parts correspond to the non-depleted parts of a 2DEG buried in a Ga[Al]As heterostructure. In both samples, a QD is formed using the Schottky gates G1, G2, and G3, (upper half of the image) and is electrically separated from a QPC charge-readout circuit (lower half) by an oxide line. On sample A, the readout QPC is formed between the metal gate G4 and the oxide line, whereas on sample B the QPC is formed by a second oxide line. The gray shaded metal gates marked `X' have not been used.} \label{fig:PS_figure_Intro}
\end{figure}

\section{Experimental technique}
\label{sec:Setup}
The experiments were done on two different QD samples shown in Fig.~\ref{fig:PS_figure_Intro}. They were fabricated with a combination of electron-beam and scanning-probe lithography \cite{Roessler10} on a $\mathrm{GaAs/Al_{0.3}Ga_{0.7}As}$ heterostructure containing a two-dimensional electron gas (2DEG) $34 \,\mathrm{nm}$ below the surface (density $4.9 \times 10^{11} \, \mathrm{cm}^{-2}$, mobility $3.3 \times 10^5 \mathrm{cm^2/Vs}$ at $4.2 \, \mathrm{K}$). In both samples, negative bias voltages on the Schottky gates G1, G2, and G3 define the quantum dot with a charging energy $E_C$ of around $1 \, \mathrm{meV}$ and a typical single-particle level spacing $\Delta$ of $100 \, \mathrm{\mu eV}$. The measurements were done in a $^3$He/$^4$He dilution refrigerator with a base temperature of $\sim 10 \, \mathrm{mK}$. A bias voltage $V_\mathrm{QD}$ was applied symmectically between the source and drain leads of the QD, and the differential conductance $g_\mathrm{QD} = dI_\mathrm{QD}/dV_\mathrm{QD}$ was measured with standard lock-in technique. The charge signal of the QPC was measured via the transconductance $g_\mathrm{QPC-TC} = dI_\mathrm{QPC}/dV_\mathrm{G2}$ at a second lock-in frequency. To this end, the QPC was biased with a constant source--drain voltage of 500 to $700 \, \mathrm{\mu V}$, and the voltage on the QD gate G2 was modulated with small amplitude ($100 \, \mathrm{\mu Vrms}$ or less) \cite{Sprinzak02}. Lock-in integration time constants ranging from 0.3 to $10 \, \mathrm{s}$ have been used. In order to optimize the signal strength of the charge detector, the QPC conductance was tuned to a value of around $0.4\times 2e^2/h$ using gate G4.

\section{Thermally and lifetime-broadened lineshapes}
\label{sec:LoHiCoupling}

\begin{figure}
\includegraphics{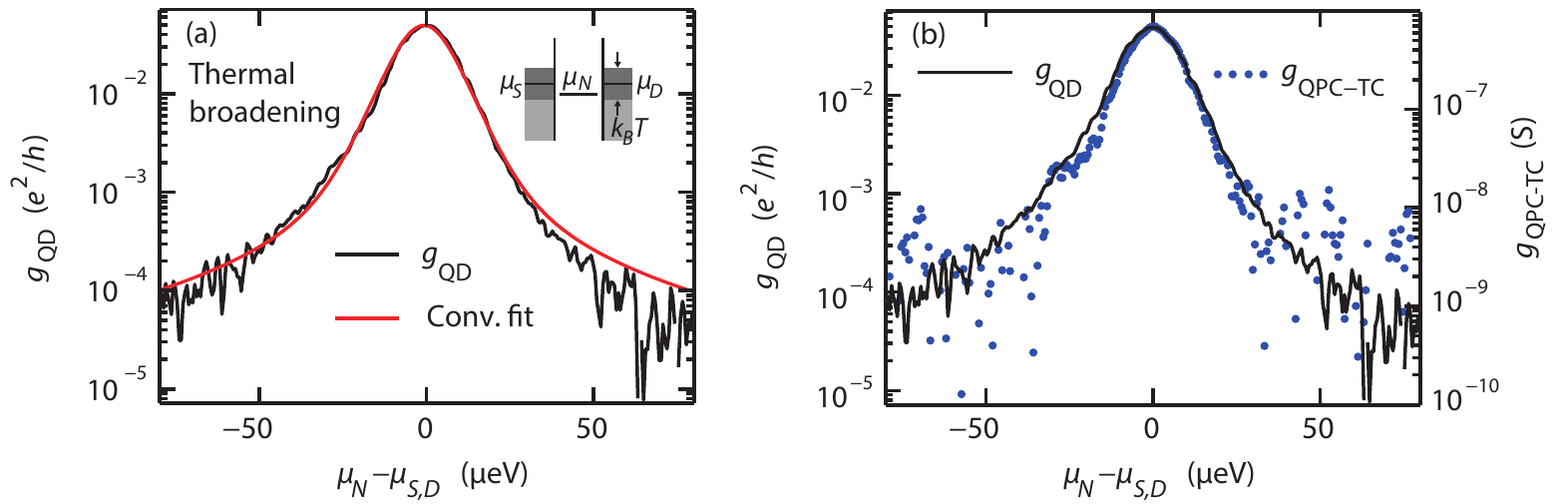}
\caption{(a) Differential QD conductance $g_\mathrm{QD}$ (black) in the regime of weak dot--lead coupling along with a fit (red) to a Fermi--Lorentz convolution (cf.~main text; fit parameters $k_BT = 4.4 \, \mathrm{\mu eV}$, $\hbar\Gamma = 3.6\, \mathrm{\mu eV}$). (b) Black solid curve: same data as in (a). The blue dotted curve is the transconductance signal measured with the QPC. Both data sets have been smoothened over a range of $2 \, \mathrm{\mu e V}$ (5 data points).}
\label{fig:PS_figure_lowCoupling}
\end{figure}

\begin{figure}
\includegraphics{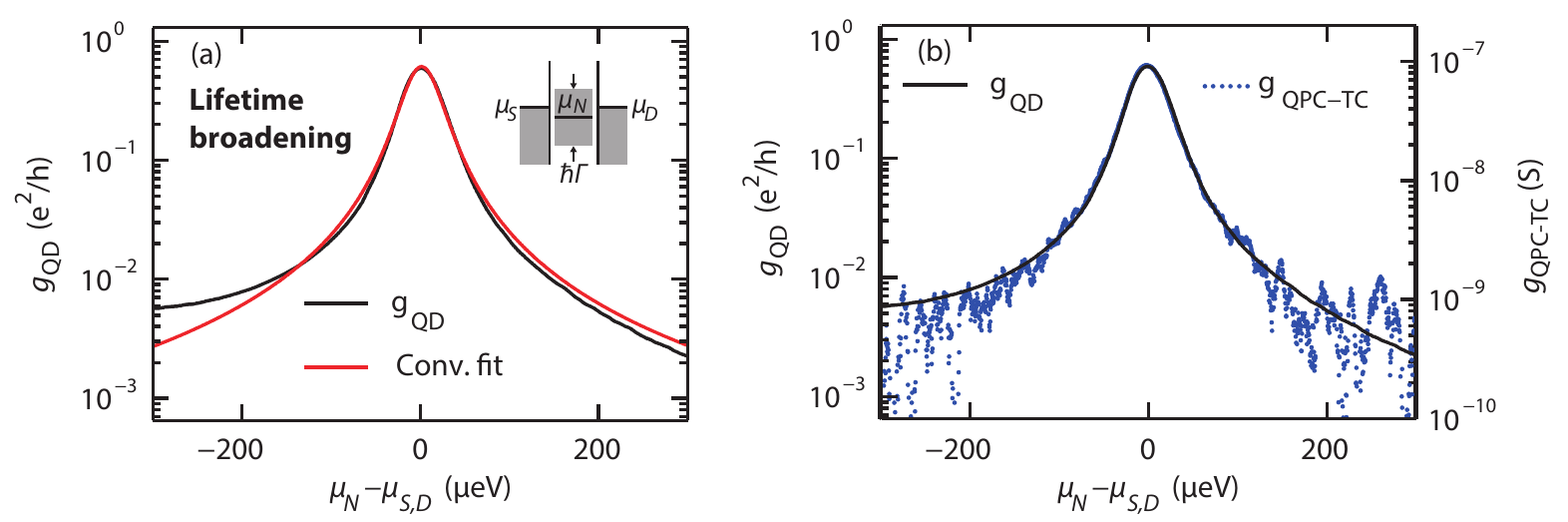}
\caption{(a) Differential QD conductance $g_\mathrm{QD}$ (black) in the regime of strong dot--lead coupling along with a fit (red) to a Fermi--Lorentz convolution (cf.~main text; fit parameters $\hbar\Gamma = 37.8 \, \mathrm{\mu eV}$, whereas the parameter $k_BT = 4.4 \, \mathrm{\mu eV}$ was held fixed). (b) Black solid curve: same data as in (a). The blue dotted curve is the transconductance signal measured with the QPC. Both data sets have been smoothened over a range of $7.5 \, \mathrm{\mu eV}$ (20 data points).} \label{fig:PS_figure_hiCoupling}
\end{figure}

The coupling strength of the QD state to the leads, $\Gamma_S$ to source, and $\Gamma_D$ to drain, is controlled with the gates G1 and G3. The lifetime broadening of the QD state $\hbar \Gamma = \hbar (\Gamma_S+\Gamma_D)$ can be continuously tuned from below to above the thermal energy $k_BT$ corresponding to the temperature of the electrons in the leads \cite{Foxman93}. The larger of the two energy scales determines the line width of the conductance resonances of the QD. The limiting regimes are characterized as thermally activated single-level transport in case $\hbar \Gamma \ll k_BT \ll \Delta$, and as a Breit--Wigner transmission resonance in case $k_BT \ll \hbar \Gamma \ll \Delta$. We experimentally approach the two regimes in the measurements shown in Figs.~\ref{fig:PS_figure_lowCoupling} and \ref{fig:PS_figure_hiCoupling}. In the first case, the shape of a peak in the QD conductance at zero $V_\mathrm{QD}$ is given by \cite{Beenakker91}
\begin{equation}
\label{eq:Fermi_formula}
g_\mathrm{QD}(E) = \frac{e^2}{4k_BT}\frac{\Gamma_S\Gamma_D}{\Gamma_S+\Gamma_D} \frac{1}{\cosh^2(E/2k_BT)},
\end{equation}
where $E = \mu_N-\mu_S =  \mu_N-\mu_D $ is the difference between the electrochemical potential of the nondegenerate QD level ($\mu_N$) and that of the leads ($\mu_S$, $\mu_D$). In the second case, tunneling through the QD is well described as a double-barrier scattering process of independent particles, and the conductance peak takes the Lorentzian form of the corresponding transmission probability,
\begin{equation}
\label{eq:Lorentz_formula}
g_\mathrm{QD}(E) = \frac{e^2}{h}\frac{\Gamma_S\Gamma_D}{\Gamma_S+\Gamma_D}\frac{\Gamma}{(E/\hbar)^2 + (\Gamma/2)^2}.
\end{equation}
When increasing the dot--lead coupling further, Kondo correlations emerge which render the single-particle approximation and equation (\ref{eq:Lorentz_formula}) invalid. In the measurements presented here, the coupling strength was kept well below values at which these correlations are typically observed.

In order to describe the line shape of $g_\mathrm{QD}$ in the intermediate regime, where the energy scales $\hbar\Gamma$ and $k_B T$ are comparable, it is often a good approach to use a convolution \cite{Foxman93} of equation (\ref{eq:Lorentz_formula}) with the energy derivative of a Fermi--Dirac distribution function. In Fig.~\ref{fig:PS_figure_lowCoupling}(a), we plot a QD-conductance peak in the weak-coupling regime, along with a fit to such a convolution (fitting parameters $\hbar\Gamma = 3.6 \,\mathrm{\mu eV}$ and $k_BT = 4.4 \,\mathrm{\mu eV}$ corresponding to a temperature of about $50 \, \mathrm{mK}$). Figure \ref{fig:PS_figure_hiCoupling}(a) shows a measurement at stronger coupling, where the lifetime broadening of the conductance peak exceeds thermal broadening (fitting parameter $\hbar\Gamma = 38 \,\mathrm{\mu eV}$, while $k_BT = 4.4 \,\mathrm{\mu eV}$ was fixed to the value found in the weak-coupling case). The small asymmetry with respect to $\mu_{N}-\mu_\mathrm{S,D} = 0 \, \mathrm{\mu eV}$ is mainly caused by the overlap with the next Coulomb blockade peak towards higher gate voltages.

In both regimes, the QPC transconductance signal was measured simultaneously with the QD conductance. In Figs.~\ref{fig:PS_figure_lowCoupling}(b) and \ref{fig:PS_figure_hiCoupling}(b), we plot the two signals on top of each other for comparison. The scaling of the vertical axes is chosen such to achieve an optimal overlap of the curves. Indeed, QD conductance and QPC transconductance match well over the covered range of signal strength, about two orders of magnitude.

As the measurements are done at zero (direct) source--drain voltage, the QD is in thermal equilibrium with its leads. In the idealized weak-coupling case, the QD level has negligible width compared to $k_BT$. The time-averaged occupation number $n(E)$ of the QD is then determined by the Fermi--Dirac distribution of the electrons in the leads, $n(E)= N+1/[1+\exp(E/k_BT)]$, up to an integer offset $N$ of electrons on the dot. Assuming a constant gate lever arm $dE/dV_\mathrm{G2}$, the QPC signal is then given by $g_\mathrm{QPC-TC} = \Delta I_\mathrm{QPC} n'(E) dE/dV_\mathrm{G2}$, where $\Delta I_\mathrm{QPC}$ is the sensitivity of the QPC current to a QD occupation change of 1 electron. The derivative $n'(E)$, and thus the QPC signal, is proportional to equation (\ref{eq:Fermi_formula}) \cite{Averin91}. In the strong-coupling case, the occupation is determined by the spectral density of the QD state which is Lorentzian, and thus the QPC signal is also expected to exhibit the same line shape as the QD conductance \cite{Buettiker93b,Pretre96}. We note that our single-level transport situation is different from that of a multi-level or even metallic dot. In that case, neither equation (\ref{eq:Fermi_formula}), nor equation (\ref{eq:Lorentz_formula}), nor their convolution accurately  describes the measurement of the QD charge as demonstrated in Ref.~\cite{Berman99}. We tested the validity of the single-level transport assumption by measuring the temperature dependence of our data (not shown). 

\section{Inelastic cotunneling}
\label{sec:Inelastic_Cotunneling}
\begin{figure}
\includegraphics{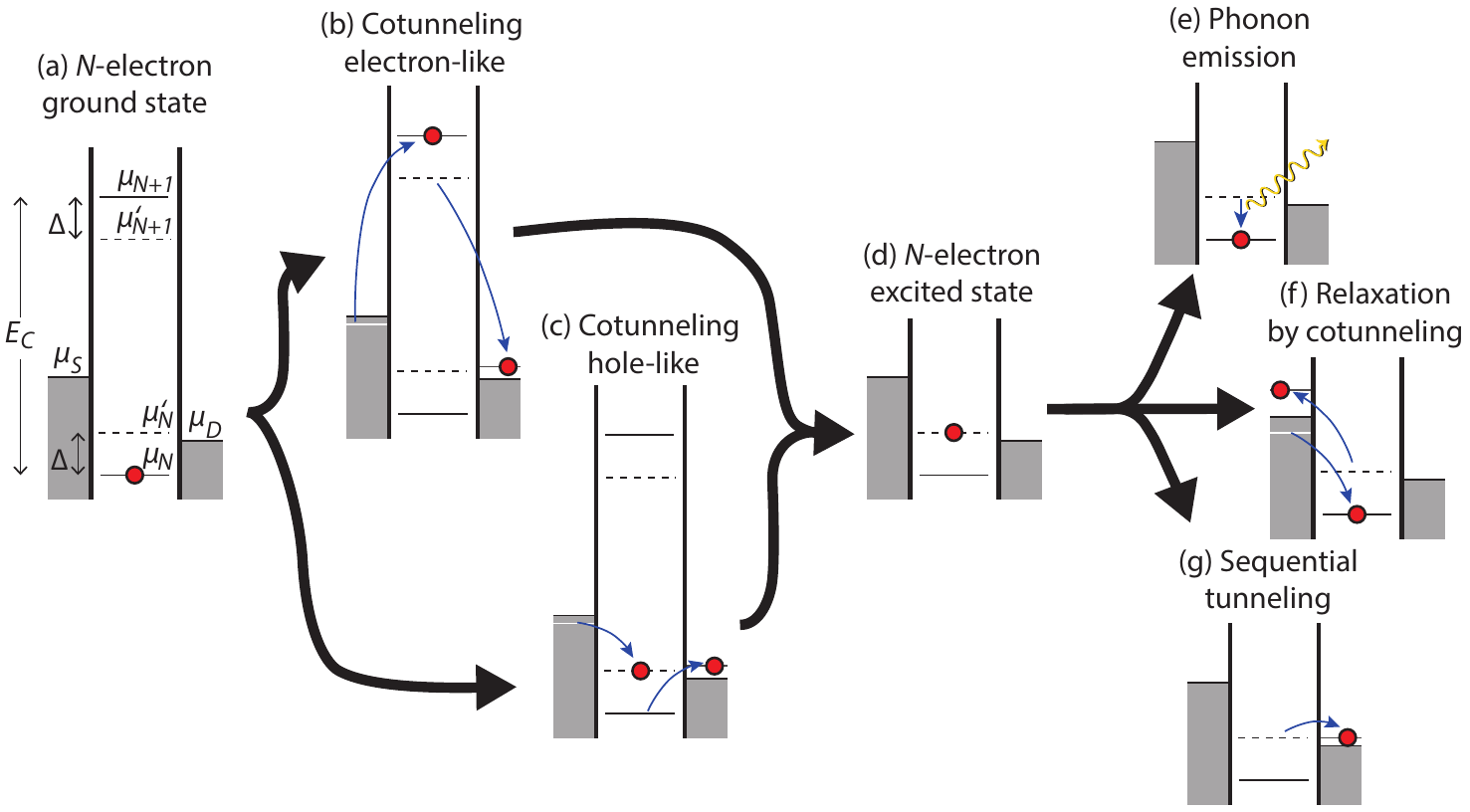}
\caption{Energy diagrams showing partial processes that can occur in the inelastic-cotunneling regime. (a) $N$-electron ground state configuration and nomenclature. (b,c) If the QD source--drain voltage $V_\mathrm{QD} = (\mu_S-\mu_D)/e$ exceeds $\Delta/e$, the QD can be brought to an excited state in a cotunneling process transferring one electron from source to drain. Two cotunneling channels contribute. The electron-like channel (b) consists of tunneling from the source into the dot followed by tunneling from the dot to the drain. The virtual intermediate state is an $(N+1)$-electron state. The blockade energy for this processe is at least $\mu_{N+1}-\mu_{S}$. The hole-like channel (c) consists of tunneling from the dot to the drain followed by tunneling from the source into the dot. The virtual intermediate state is an $(N-1)$-electron state. The blockade energy for this process is at least $\mu_{D}-\mu_{N}$. (d) After cotunneling, the dot is left in an $N$-electron excited state. Subsequently, the QD can relax by emission of a phonon or photon (e), or in a cotunneling process involving a single lead (f). In the sketch, the excited-state energy $\mu_N'$ lies above the drain level $\mu_D$, in which case the electron can tunnel elastically to the drain (g).} \label{fig:PS_figure_Sketches}
\end{figure}

Conduction of a QD in the tails of a lifetime-broadened peak is due to elastic cotunneling, a second-order process already present close to equilibrium. As the QD is driven out of equilibrium by a source--drain voltage exceeding the energy $\Delta$ of the first excited QD state, additional processes come into play. These are called inelastic cotunneling processes and bear a close analogy to their elastic counterparts. During such a process, as depicted in figure \ref{fig:PS_figure_Sketches}(a--c), an electron tunnels out to the drain from the ground state, while a second electron tunnels into the excited state from the source. The QD spends a short time in a virtual intermediate state whose energy lies outside of the classically allowed range. The sequence of the partial tunneling processes can be interchanged, which gives rise to two channels (hole-like and electron-like) which both contribute to the total cotunneling amplitude. In the electron-like sequence, figure \ref{fig:PS_figure_Sketches}(b), tunneling from the source into the dot occurs first, and the virtual intermediate state is an $(N+1)$-electron state. In the hole-like sequence, figure \ref{fig:PS_figure_Sketches}(c), tunneling from the dot to the drain occurs first, and the virtual intermediate state is an $(N-1)$-electron state. The classical blockade energies of the two channels are different in general, as is specified in the figure caption. 

\begin{figure}
\includegraphics{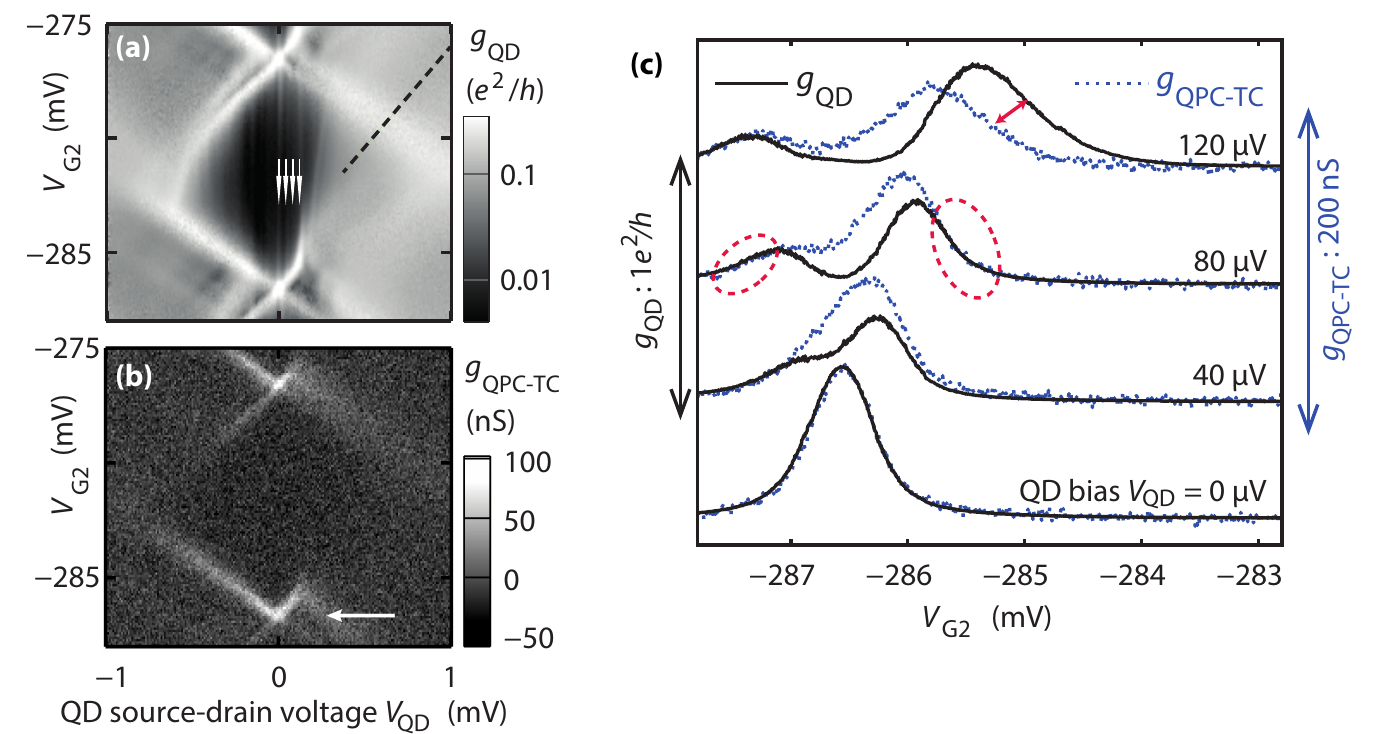}
\caption{(a) Coulomb blockade diamond measurement of $g_\mathrm{QD}$. An onset of inelastic cotunneling is visible at $V_\mathrm{QD} = 120 \, \mathrm{\mu V}$. (b) Simultaneous measurement of $g_\mathrm{QPC-TC}$. In these data, no inealstic-cotunneling onset is visible, but instead a signature (white arrow) of the excited dot state responsible for inelastic cotunneling. (c) QPC and QD signals measured along $V_\mathrm{G2}$ at four different $V_\mathrm{QD}$ indicated by arrows in (a). The two signals at zero bias are scaled to match; this scaling is maintained for the rest of the curves, which are also vertically offset for clarity. For finite bias voltages $40 \, \mathrm{\mu V}$ and $80 \, \mathrm{\mu V}$, the two signals still agree in the tails as indicated by the dashed ellipses. Beyond the inelastic cotunneling onset (at $V_\mathrm{QD} = 120 \, \mathrm{\mu V}$), the signals clearly deviate. 
} \label{fig:PS_figure_inelCotunneling}
\end{figure}

The overall transmission probability from source to drain, and thus the cotunneling current, depends on the detuning of the QD potential with respect to the leads, but the minimum source--drain voltage is independent of detuning and given by $\Delta/e$. The experimental signature of inelastic cotunneling is hence a conductance step inside a Coulomb blockade diamond parallel to the line of zero bias \cite{Franceschi01,Wegewijs01}. Such a feature is seen in the $g_\mathrm{QD}$ data shown in figure \ref{fig:PS_figure_inelCotunneling}(a) at a positive source--drain voltage of around $120 \, \mathrm{\mu V}$. In the simultaneous measurement of the QPC transconductance shown in panel (b), no finite-bias feature is visible.

The discrepancy between the two signals becomes more evident when looking at the measurement in figure \ref{fig:PS_figure_inelCotunneling}(c) taken along cuts at four source--drain voltages, as indicated by the white arrows in panel (a). At $V_\mathrm{QD} = 0 \, \mathrm{\mu V}$, the peak in the QD conductance is due to elastic cotunneling as discussed in Sec.~\ref{sec:LoHiCoupling}, with a broadening of $\hbar \Gamma = 49 \, \mathrm{\mu eV}$. As expected, the two signals $g_\mathrm{QD}$ and $g_\mathrm{QPC-TC}$ can be made to fit by scaling. Upon increasing $V_\mathrm{QD}$, both the peaks in $g_\mathrm{QD}$ and in $g_\mathrm{QPC-TC}$ broaden and eventually split. It is then not expected that the two signals match over the whole range in gate voltage. Namely, the relative height of the two sub-peaks in $g_\mathrm{QPC-TC}$ can take any value depending on the coupling symmetry \cite{Schleser05,Rogge05} $\Gamma_S/\Gamma_D$. In contrast, the two sub-peaks in $g_\mathrm{QD}$ are expected to be equal in height independent of symmetry in case of single-level transport with energy-independent coupling. The difference in height of the $g_\mathrm{QD}$ peaks in our data can be explained by energy-dependent tunneling rates $\Gamma_{S,D}$. Despite the overall discrepancies between $g_\mathrm{QD}$ and $g_\mathrm{QPC-TC}$ signals, by tuning the coupling asymmetry we can achieve that they agree in the tails (after vertical scaling) as indicated by the dashed ellipses in figure \ref{fig:PS_figure_inelCotunneling}(c). This is expected because in the limit $\mu_N-\mu_S,\, \mu_N-\mu_D \gg eV_\mathrm{QD}$, the zero-bias results for $g_\mathrm{QD}$ and $g_\mathrm{QPC-TC}$ are valid, which implies that both quantities decay proportionally to $(\mu_N-\mu_S)^{-2} \sim  (\mu_N-\mu_D)^{-2}$.

As the bias is increased above the energy of the excited state, $V_\mathrm{QD} = 120 \, \mathrm{\mu V}$, the signals clearly deviate in the right-hand tail, as indicated by the double arrow in figure \ref{fig:PS_figure_inelCotunneling}(c). At this point, the QD conductance is clearly enhanced compared to the occupation signal and its maximum shifts to the right into the Coulomb-blockaded region. This indicates that the extra conductance is not due to sequential tunneling through the excited state: in that case, we would instead expect a conductance feature shifting to the left into the bias window. We therefore assume that the extra conductance is primarily due to inelastic cotunneling.

The qualitative difference between the data at $V_\mathrm{QD} = 0 \, \mathrm{\mu V}$ and $V_\mathrm{QD} = 120 \, \mathrm{\mu V}$ demonstrates the contrasting character of the measured QD charge in the two cotunneling regimes. At zero QD voltage, the charge is entirely characterized as an equilibrium quantity. It is determined by the spectral density of the QD state and does not explicitly depend on the presence of two leads, i.e., does not depend on transport. The proportionality of the QPC signal to the QD conductance observed at low $V_\mathrm{QD}$ is a consequence of the fact that both are determined by the spectral density of the QD state. This simple relation is lost when going out of equilibrium, in particular in the regime of inelastic cotunneling. But in this regime, we can identify a different connection between the QD charge state and transport. Namely, the only property distinguishing a QD slightly below the inelastic-cotunneling onset from the same QD slightly above the onset is the fact that in the latter case, processes of the kind shown in figure \ref{fig:PS_figure_Sketches}(b,c) are energetically allowed. Because an electron is transferred through the QD in each such process, any change in the QD charge occurring at the position of the inelastic-cotunneling onset in $g_\mathrm{QD}$ is most likely linked to transport.

In order to be more quantitative, we aim at linking the direct QD current, $I_\mathrm{QD}$, to the observed QD charge state. We define the transport occupation of the QD as the frequency of carriers passing the QD multiplied by the average dwell time of each carrier on the QD, $n_\mathrm{transport} = \pm I_\mathrm{QD} \tau_\mathrm{dwell}/e$. The sign depends on whether the current is carried by electrons or holes. In the general case, the current $I_\mathrm{QD}$ would have to be split into an electron and a hole current, and the dwell times of the two carrier types could be different. This corresponds to the two parallel cotunneling channels discussed in the beginning of this section. We will only look at cases where either of the two carrier types dominates, so we can ignore this complication.

The transport occupation $n_\mathrm{transport}$ is in general not equal to the total occupation $n$ which is relevant for the measurement. The two are equal under the conditions that the current is constituted of a single type of tunneling processes (all with a similar dwell time), and that tunneling processes are absent which contribute to the occupation but not to the current. In case these conditions are met, the measured QPC signal can be expressed as
\begin{equation}
\label{eq:QPC-TC_calculated}
g_\mathrm{QPC-TC}^\mathrm{calc} = \frac{d(\tau_\mathrm{dwell}I_\mathrm{QD}/e)}{dV_\mathrm{G2}}\times\Delta I_\mathrm{QPC},
\end{equation}
where $\Delta I_\mathrm{QPC}$ is the sensitivity of the QPC current to an occupation change of one electron.

\begin{figure}
\includegraphics{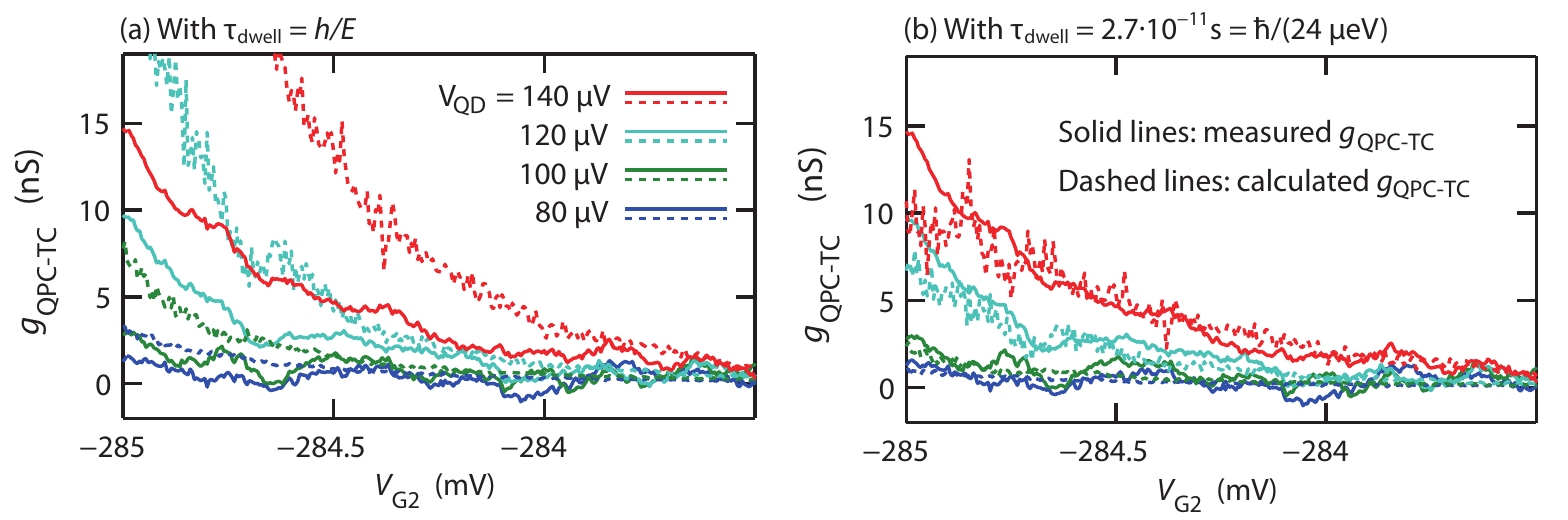}
\caption{Comparison of measured with calculated QPC signal in the inelastic-cotunneling regime. Solid lines in (identical in a and b) show $g_\mathrm{QPC-TC}$ measured at different QD source--drain voltages $V_{QD}$ in the configuration of figure \ref{fig:PS_figure_inelCotunneling}. The equilibrium contribution to $g_\mathrm{QPC-TC}$ at $V_\mathrm{QD}=0 \, \mathrm{\mu V}$ has been subtracted from all curves. The dashed lines in (a) show the theoretical QPC transconductance signal calculated from $I_\mathrm{QD}$ and a particle dwell time $\tau_\mathrm{dwell} = h/(\mu_D-\mu_N)$, equation (\ref{eq:QPC-TC_calculated}). The dashed lines in (b) show the corresponding calculation when assuming an energy-independent dwell time $\tau_\mathrm{dwell} = 2.7 \times 10^{-11}\, \mathrm{s}$. All data have been smoothened over a range of $100 \, \mathrm{\mu V}$ in $V_\mathrm{G2}$.} \label{fig:PS_figure_DirectCurrentAndTC}
\end{figure}

Figure \ref{fig:PS_figure_DirectCurrentAndTC} shows charge sensing data for $V_\mathrm{QD}$ values around the inelastic-cotunneling onset. They belong to the same measurement as those in figure \ref{fig:PS_figure_inelCotunneling}(c). But these traces have a smaller $V_\mathrm{G2}$ range restricted to the region in which the inelastic-cotunneling contributions to $g_\mathrm{QD}$ and $g_\mathrm{QPC-TC}$ (at bias voltages just above $V_\mathrm{QD}=120 \, \mathrm{\mu V}$) clearly dominate over the remaining elastic contributions (just below $V_\mathrm{QD}=120 \, \mathrm{\mu V}$). The solid traces show the measured $g_\mathrm{QPC-TC}$ minus a small contribution at $V_\mathrm{QD} = 0 \, \mathrm{\mu V}$.

In the case of an inelastic cotunneling process, the dwell time corresponds to the lifetime of the intermediate virtual QD state. Near the cotunneling onset, its energy lies outside the classically allowed range by an amount $E=\mu_D-\mu_N$. The lifetime of a virtual state in a setup like this is often treated using an energy-time uncertainty relation in the spirit of Heisenberg \cite{Heisenberg27}. The heuristic relation $\Delta E \times\Delta t \sim h$ involves the quantum uncertainty $\Delta E$ of the energy of a state during a process and the duration $\Delta t$ of that process. In the context of cotunneling, it means that if an electron resides in the dot for a short time $\Delta t \lesssim h/E$, its energy is necessarily dispersed enough that the difference $E$ between dot and lead state energies becomes irrelevant and the amplitude for a tunneling process through the dot becomes nonzero. The intermediate state has a dispersed energy because it is a superposition between two states with rather sharp energies 0 and $E$. A dwell time on the dot exceeding $h/E$ on the other hand would mean that the energy of the intermediate state is well-defined at the value $\mu_N$, which would violate energy conservation because the initial and final state energy must lie within the bias window. We note, however, that the above energy-time relation is not strictly a version of the usual Heisenberg uncertainty principle of noncommuting operators, such as position and momentum. This is because time, unlike energy, is not a quantum operator.

To compare our data with this theoretical picture, we insert the value $\tau_\mathrm{dwell}=h/E$ into equation (\ref{eq:QPC-TC_calculated}) to calculate $g_\mathrm{QPC-TC}^\mathrm{calc}$ which we compare with the measurement. The dashed lines in figure \ref{fig:PS_figure_DirectCurrentAndTC}(a) show the result of equation (\ref{eq:QPC-TC_calculated}). Indeed, measured and calculated charge signals are of similar magnitude, and apart from noise the measured signal is always smaller than the calculation, indicating a dwell time bounded by $h/E$.

Note that in case of a \emph{resonant} process (on-peak), instead one typically assumes a constant dwell time $\tau_\mathrm{dwell} = 1/\Gamma$ equal to the lifetime of the QD state \cite{Buks98}. The cotunneling picture then loses its validity, which can also be recognized from the fact that the time $h/E$ diverges when $E$ tend to zero. Namely, we have justified this expression for $\tau_\mathrm{dwell}$ by recognizing that the energy uncertainty of the quantum state during tunneling is determined by the blockade energy $E$. If $E$ tends to zero, the energy uncertainty will eventually be limited by the intrinsic width $\hbar \Gamma$ of the dot states participating in tunneling, and we enter the regime of resonant tunneling. 

Assuming a constant dwell time, as is done in the case of resonant tunneling, is problematic in the case of cotunnrling however, since a constant dwell time violates the Heisenberg relation off-peak. Nevertheless, it is possible to achieve a good description of the transconductance data by assuming a constant dwell time as seen in figure \ref{fig:PS_figure_DirectCurrentAndTC}(b). The value of $\tau_\mathrm{dwell} = 2.7 \times 10^{-11} \, \mathrm{s} = \hbar/(24 \, \mathrm{\mu eV})$ we used there was chosen such to achieve a good fit of the data. There is no further physical justification for it, but the plots may help to judge about the comparison of measurement and calculation in the left column. The fact that the data can be reproduced using a constant dwell time means that the main energy dependence of $g_\mathrm{QPC-TC}$ comes from the energy dependence of $I_\mathrm{QD}$. It is generally not surprising that in the constant-time case, data and calculation agree better than in the Heisenberg time case. When assuming a constant dwell time, there is one free parameter (the constant value of $\Gamma$), but there is none when assuming $\tau_\mathrm{dwell}=h/E$. The comparison between the two cases tells us that the Heisenberg time $h/E$ is probably not a good approximation for the actual dwell time, but it appears that its role as an upper limit to the dwell time is relevant for the measurement.

The validity of the picture leading to equation (\ref{eq:QPC-TC_calculated}) depends, as mentioned, on the additional processes taking place after the initial inelastic cotunneling event. Due to the variety of conceivable tunneling and relaxation processes at large QD bias, it is necessary to discuss their possible influence on the data in figures \ref{fig:PS_figure_DirectCurrentAndTC} and \ref{fig:PS_figure_inelCotunneling}. Figure \ref{fig:PS_figure_Sketches} (e) through (g) shows the three main options for the continuation of the QD evolution after an inelastic cotunneling event (panel a) has brought it to the excited state. All these options eventually bring the QD back to the ground state and thus close the transport cycle. In the simplest case sketched in panel (c), the electron relaxes to the ground state through emission of a phonon \cite{Fujisawa02}, a process which neither contributes to the QD current nor changes its charge state. A typical phonon emission rate \cite{Fujisawa02} in GaAs QDs is $0.1 \, \mathrm{GHz}$. Alternatively, the QD may relax through another inelastic cotunneling process involving two, or also just one lead as shown in figure \ref{fig:PS_figure_Sketches}(f) for the case of the source lead. All these processes contribute to the occupation of the QD. Finally, sequential tunneling to the drain lead may occur if the excited-state energy $\mu_N'$ lies above the drain potential $\mu_D$. In the absence of a pronounced feature in our data at the zero-crossing of $\mu_N'-\mu_D$ (the continuation of the dashed line in figure \ref{fig:PS_figure_inelCotunneling}[a]), we conclude that such processes don't contribute significantly to either occupation or current (possibly due to a poor coupling of the excited state to the drain).

What remains are thus relaxation processes as shown in panel (d) which compete with phonon emission. An estimate of the inelastic cotunneling relaxation rate \cite{Fujisawa02} requires the knowledge of the coupling of both ground and excited state to the source reservoir. While the ground state coupling of around $20 \, \mathrm{GHz}$ can be reliably inferred from finite-bias charge sensing \cite{Schleser05}, determining the excited-state coupling is more difficult. From the clear charging feature at the entrance of the excited state in the bias window (arrow in figure \ref{fig:PS_figure_inelCotunneling}[b]), we conclude that the excited-state coupling must be at least comparable to the ground-state coupling. This would lead to an inelastic-cotunneling relaxation rate of at least 6 to $20 \, \mathrm{GHz}$ in the gate voltage range of figure \ref{fig:PS_figure_DirectCurrentAndTC}, which thus dominates over phonon emission. The consequence is an effective dwell time per transport cycle which includes the dwell time of the hole during the relaxation process, on top of the dwell time for the initial cotunneling process. Since the blockade energy $E$ is the same for both processes, the bound $h/E$ on the dwell times is identical.

\section{Cotunneling-assisted sequential tunneling}
The character of the transport process can be changed strongly in case that not the relaxation processes of the kind shown in figure \ref{fig:PS_figure_Sketches}(e,f), but instead the sequential process shown in figure \ref{fig:PS_figure_Sketches}(g) is dominant. Such a sequential process leaves the QD in a state with $N-1$ electrons. Another electron can then tunnel elastically from source into the excited state, and this sequential cycle may be repeated several times, until relaxation or tunneling into the $N$-electron ground state takes place. This effect is called cotunneling-assisted sequential tunneling (CAST) \cite{Wegewijs01,Golovach04,Schleser05cast,Aghassi08}.
 
\begin{figure}
\includegraphics{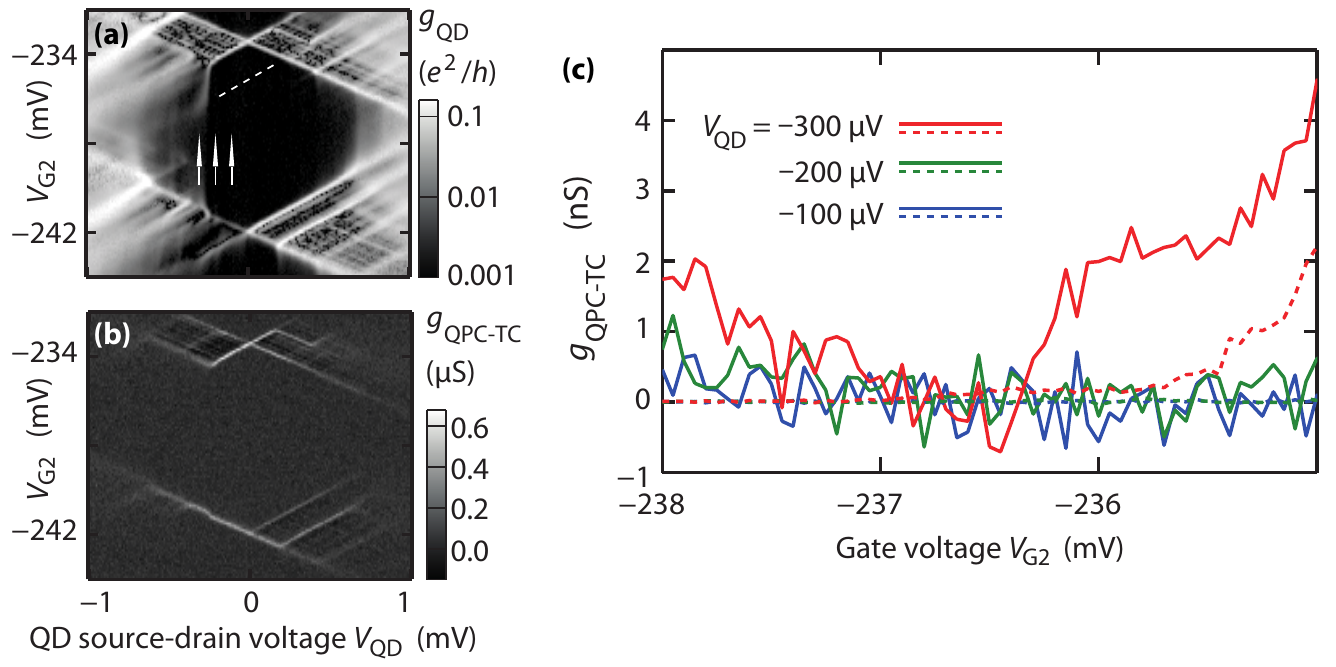}
\caption{(a) Coulomb blockade diamond measurement of $g_\mathrm{QD}$. An onset of inelastic cotunneling is visible at $V_\mathrm{QD} = -230 \, \mathrm{\mu V}$. In addition, a signature of CAST is visible as the extension of the dashed line towards negative $V_\mathrm{QD}$. (b) Simultaneous measurement of $g_\mathrm{QPC-TC}$. (c) Solid lines: QPC signal $g_\mathrm{QPC-TC}$ measured along $V_\mathrm{G2}$ at three different $V_\mathrm{QD}$ indicated by arrows in (a). The signal was integrated longer than in the diamond measurement to reveal the faint CAST signature in the measurement at $V_\mathrm{QD} = -300 \, \mathrm{\mu V}$. Dashed lines: Theoretical QPC transconductance signal calculated from $I_\mathrm{QD}$ and the particle dwell time $\tau_\mathrm{dwell} = h/(\mu_\mathrm{QD}-\mu_\mathrm{source})$, equation (\ref{eq:QPC-TC_calculated}). The measured signal clearly exceeds the calculated one due to the long dwell times of the sequential tunneling processes occurring in this regime. This is in contrast to the case of pure inelastic cotunneling (figure \ref{fig:PS_figure_DirectCurrentAndTC}). Smoothened over a range of $150 \, \mathrm{\mu V}$ in $V_\mathrm{G2}$.
\label{fig:PS_figure_InCo2}}
\end{figure}

There is a clear experimental signature for CAST: a conductance step inside a Coulomb blockade diamond parallel to the diamond edge \cite{Schleser05cast}. This line borders the region in which CAST is energetically allowed. The conditions to observe such a feature are special. The QD needs to be in a configuration featuring an excited state that is more strongly coupled to source and drain than the ground state, but weakly enough to maintain sufficient energy resolution. Furthermore, relaxation from the excited to the ground state needs to be slow. Albeit only partially under experimental control, these conditions vary randomly from one Coulomb-blockade diamond to the next due to the quantum nature of the dot states. This allows for a systematic search of CAST features in measurement. In the Coulomb-blockade diamond shown in figure \ref{fig:PS_figure_InCo2}(a), such a feature is visible as the continuation of the dashed line towards negative $V_\mathrm{QD}$. Neither inelastic-cotunneling nor CAST onsets are visible in the simultaneous measurement of $g_\mathrm{QPC-TC}$ in panel (b). More accurate line cuts in figure \ref{fig:PS_figure_InCo2}(c) reveal, however, that $g_\mathrm{QPC-TC}$ is non-zero in the cotunneling regime. Above the inelastic cotunneling onset at $V_\mathrm{QD} = -300 \, \mathrm{\mu V}$, a QPC signal of the order of $2 \, \mathrm{nS}$ builds up inside the region where CAST is allowed (for $V_\mathrm{G2} > -236.1 \, \mathrm{mV}$).

The particle dwell time in a sequential process is associated with a real, rather than virtual, intermediate state and as such is not limited by the Heisenberg relation as it was for a cotunneling process. To emphasize this difference, we plot as dashed lines in figure \ref{fig:PS_figure_InCo2}(c) the result of equation (\ref{eq:QPC-TC_calculated}), taking into account the direct QD current and the cotunneling dwell time $h/E$. The calculated trace at a bias above the cotunneling onset ($V_\mathrm{QD} = - 300 \, \mathrm{\mu V}$) exhibits a clear enhancement of the expected charge signal compared to the traces below the onset. This is a qualitative feature shared with the measured trace. Quantitatively the two disagree like in the case of pure cotunneling, but unlike there the measured signal is stronger than the calculated one, signaling a dwell time exceeding $h/E$.

\section{Summary}
\label{sec:Summary}
In summary, we have presented simultaneous measurements of the conductance and charge occupation of a QD in several parameter regimes. At zero QD voltage, the charge signal is successfully interpreted in terms of the equilibrium occupation and the equilibrium conductance of the QD. Both in the regimes of weak coupling (dominated by thermal broadening) and of strong coupling (dominated by lifetime broadening), conductance and charge signal lineshapes are found to agree. At nonzero QD voltage, we study the inelastic cotunneling regime where the charge signal is generally weaker, and where an analysis in terms of equilibrium occupation is not possible. 

We compare the charge signal to a theoretical signal calculated from the QD current and a charge carrier dwell time of $h/E$, where $E$ is the blockade energy of first-order tunneling. Such a dwell time estimate is often derived from a cotunneling picture involving a virtual intermediate state with a lifetime bounded by an energy-time uncertainty principle. Since our measured charge occupation is smaller than the calculated one, our results support this cotunneling picture experimentally. Assuming an energy-independent dwell time allows for a good fit of the data in the accessible range of energies $E$. 

Further measurements carried out in a regime of mixed sequential tunneling and inelastic cotunneling are consistent with the above interpretation. The charge occupation signal observed in this regime clearly exceeds the calculation based on a dwell time of $h/E$. This is expected because of the longer carrier dwell time in sequential tunneling as compared to that in cotunneling.

\section*{Acknowledgements}
We thank C. Beenakker, K. Kobayashi, L. Levitov, M. Sanquer, G. Sch\"on, and S.E. Ulloa for discussions. We also thank M. B\"uttiker for clarifying the role of the time uncertainty in a tunneling process, and Y. Gefen for insightful discussions from the point of view of weak measurements. Sample growth and processing was mainly carried out at FIRST laboratory, ETH Zurich. Financial support from the Swiss National Science Foundation (Schweizerischer Nationalfonds, National Center for Competence in Research, Quantum Science and Technology) is gratefully acknowledged.

\end{document}